# Controllable distant interactions at bound state in the continuum

Haijun Tang[1, †], Can Huang[1, †, #], Yuhan Wang[1], Xiong Jiang[1],
Shumin Xiao[1,2,3,4], Jiecai Han[2], Qinghai Song[1,3,4,*]

1. Ministry of Industry and Information Technology Key Lab of Micro-Nano Optoelectronic Information System, Guangdong Provincial Key Laboratory of Semiconductor Optoelectronic Materials and Intelligent Photonic Systems, Harbin Institute of Technology, Shenzhen 518055, P. R. China.
2. National Key Laboratory of Science and Technology on Advanced Composites in Special Environments, Harbin Institute of Technology, Harbin 150080, P. R. China.
3. Pengcheng Laboratory, Shenzhen 518055, P. R. China.
4. Collaborative Innovation Center of Extreme Optics, Shanxi University, Taiyuan 030006, Shanxi, P. R. China.

† These authors contribute equally to this work.

Corresponding authors:
* qinghai.song@hit.edu.cn; # huangcan@hit.edu.cn;

## Abstract:

**Distant interactions at arbitrary locations and their dynamic control are fundamentally important for realizing large-scale photonic and quantum circuits. Conventional approaches suffer from short coupling distance, poor controllability, fixed locations and low wavelength uniformity, significantly restricting the scalability of photonic and quantum networks. Here, we exploit the intrinsic advantages of optical bound state in the continuum (BIC) and demonstrate an all-in-one solution for dynamically controllable long-range interactions. BIC metasurface can support a series of finite-sized quasi-BIC microlasers at arbitrary locations. The quasi-BICs microlasers have the same wavelength and are inherently connected through BIC waveguide. Consequently, the coupling distances in experiment increase significantly from subwavelength to tens of micrometers. Such long-range interaction in BIC metasurface enables scaling to two-dimensional architectures and ultrafast control of internal laser actions, e.g., non-Hermitian zero-mode lasing and enhanced optical gain. This research shall facilitate the advancement of scalable and reconfigurable photonic networks.**

Coupled arrays of advanced photon sources are critical to advances in optical computing and quantum information processing.[1-7] Such arrays usually consist of a large number of micro- or nano-sized cavities linked with their mutual coupling. Over the past decades, there has been rapid progress in individual photonic devices. High quality (Q) factors, small effective mode volume, and the corresponding light-mater interactions have been intensively explored in both classical and quantum regimes.[8-10] Nonetheless, the scaling from discrete sites to large-scale photonic or quantum networks, is strongly hampered by the short interaction distance. Tight confinement of light in each resonator improves the performance of individual nanophotonic devices, but also strongly limits its coupling to adjacent nodes within a fraction of wavelength[5-7]. Several recent techniques utilizing zero-refractive index materials[11], hyperbolic metamaterial[12], and Weyl point in engineered nanostructures[13] have demonstrated the theoretical potential to improve the coupling distance and its dynamical control. Practically, they still face the severe challenge of the trade-off between coupling range and strength.[14] One exception is the waveguide that can extend the interaction range far beyond evanescent field without compromising the strength.[15,16] However, conventional waveguides are limited to one-dimensional configuration, and the involved resonators typically have fixed locations and inevitable wavelength detuning, severely restricting the realization of reconfigurable and scalable networks.

In searching for strategies for controllable long-range interaction, we turn to the optical bound state in the continuum (BIC).[17] BIC refers to the state that remains localized despite residing in the continuum spectrum of radiation. Optical BICs have been intensively studied in a variety of nanophotonic structures due to their extremely high quality (Q) factors and unprecedented capability in controlling the topological structures in momentum space.[18-21] In principle, BIC require the engineered nanostructures to be infinitely large in at least one dimension.[17, 22] This requirement is generally regarded as a major drawback of BIC, seriously affecting the integration density and internal light-matter interaction of BIC devices. The inherent advantages of optical BIC in long-range interactions and coupled photonic arrays, however, have been simply overlooked for a very long time. Here, we take quasi-BIC microlaser as an example to demonstrate a new platform for controllable long-range interaction and scalable photonic circuits.

**The working principle of distant coupling**

The schematic of our metasurface is depicted in Fig. 1(a). It is composed of an active membrane with a thickness of $h_1$ = 100 nm on a glass substrate. The membrane is covered with a 150 nm polymer film that is periodically patterned with square-lattice air holes. The lattice size and diameter of hole are $l$ = 333 nm and $D$ = 167 nm, respectively. For the case of infinite period, numerical simulation (see Methods and Supplementary Note-1) reveals that two resonances with transverse magnetic (TM, E is perpendicular to the plane) polarization appear within the gain spectrum of active layer (shadowed area in Fig. 1(b)). The lower branch corresponds to the well-known symmetry-protected BIC at Γ-point. The coherent destruction of scattering waves confines electromagnetic field as guiding waves in the active membrane. As a result, the Q factor increases exponentially approaching Γ-point and reaches a maximal value of $10^{10}$ (Fig. 1(c)).

The ideal BIC degrades to quasi-BIC when the metasurface is excited by a laser beam of finite size (**Supplementary Note-1**).[19] The electromagnetic waves at the BIC wavelength are amplified and lase in the pump regions. Such quasi-BIC microlasers produce vector beams in the vertical direction. Their in-plane radiations are less confined and propagate to distant places through the non-radiative BIC mode in waveguide (top panel in Fig. 1(b)). The situation becomes very intriguing when two or more pump beams are applied (bottom panel in Fig. 1(b)). Different from conventional photonic systems, the quasi-BIC microlasers are defined by the external excitations and can be generated at arbitrary locations of BIC metasurface. On one hand, the laser systems can be easily reconfigured. On the other and, their operation wavelengths are determined by the lattice size and are naturally the same. Such quasi-BIC microlasers are inherently linked with the non-radiative BIC mode. Then the trade-off between coupling range and strength is simply broken and a complex network can be constructed in the two-dimensional BIC metasurface. Therefore, BIC metasurface can be an ideal platform scalable and reconfigurable photonic circuits.

For simplicity, we consider the interaction between two quasi-BIC microlasers. When the excitation is not far above the laser threshold, the gain saturation is absent and the interaction can be expressed as

$$i\frac{da}{dt} = \omega_0 a + i(\gamma_a - \kappa_a)a + J_{ba}b, \tag{1a}$$

$$i\frac{db}{dt} = \omega_0 b + i(\gamma_b - \kappa_b)b + J_{ab}a. \tag{1b}$$

Here $a$ and $b$ are the amplitudes of two quasi-BICs with the same angular frequency of $\omega_0$. $\gamma_{a,b}$ and $\kappa_{a,b}$ represent the gain and loss coefficients, respectively. The coupling is mediated by the BIC waveguide and its coefficients are defined as $J_{ab}$ and $J_{ba}$ (here we set $J_{ab} = J_{ba} = J$). Then the eigenvalues are given as

$$\omega_\pm = \omega_0 + i(\gamma_{avg} - k) \pm \sqrt{J^2 - (\Delta\gamma)^2}, \tag{2}$$

where $\Delta\gamma = (\gamma_a - \gamma_b)/2$, $\gamma_{avg} = (\gamma_a + \gamma_b)/2$, and $k_a = k_b = k$. The entire system operates as a single quasi-BIC microlaser when $\gamma_a$ is tuned only and $\gamma_b$ is fixed as zero. The coupling occurs when the gain coefficients $\gamma_a$ is kept at above the threshold and $\gamma_b$ becomes nonzero. With the increase of $\gamma_b$, the imaginary parts of eigenvalues ($Im(\omega_\pm)$) approach one another and merge at $\Delta\gamma = J$, whereas the real parts ($Re(\omega_\pm)$) remain at their initial values and bifurcate (Figs. 1(e,f)). Then the coupled quasi-BIC microlasers can be considered as a quasi-parity-time symmetric system and reaches the exceptional point (EP) at $\Delta\gamma = J$.[23-29]

In BIC metasurface, only the waves propagating to beam B are involved in the mode interaction (**Supplementary Note-1**). The other outgoing waves only contribute to optical loss and give a relatively large loss factor $k$. As a consequence, EP can appear at relatively large $\gamma_{avg}$ and below the threshold line. One example is illustrated in Fig. 1(f). The imaginary part of the lasing mode ($Im(\omega) > 0$) decreases with the increase of gain coefficient at beam B ($\gamma_b$) and eventually both *Im(ω±)* are below the threshold line (the shadowed region). This process corresponds to the well-known phenomena of lasing self-termination and even lasing death.[30-33] Further increases of $\gamma_b$ can make the coupled system lase again (*Im(ω±) > 0*). Under such a situation, the bifurcation in real parts (*Re(ω±)*) lead to an obvious mode splitting in emission spectrum (Fig. 1(e)). Therefore, the lasing self-termination and mode splitting can be two important criteria for exploring and determining the interaction of quasi-BIC microlasers.

**Experimental demonstration of long-range coupling at BIC**

Based on the above analysis, we have fabricated the designed BIC metasurfaces on a K9 glass substrate with a standard electron-beam (E-beam) lithography process (see **Methods** and **Supplementary Note-2**).[34, 35] Periodic air holes are patterned in a 150 nm E-beam resist E-beam resist (ZEP520A) and the optical gain is provided by the underneath lead halide perovskite (MAPbBr$_3$) film with a thickness of 100 nm. The overall sample size is 100×100 μm$^2$. Figure 2(a) depicts the top-view scanning electron microscope (SEM) image of the BIC metasurface. Similar to the thicknesses, both the hole diameter and the lattice size follow the design well. Then BIC mode and the corresponding lasing actions can be expected.

The optical characteristics of BIC metasurface are explored by exciting it with a frequency doubled femtosecond laser under a home-made pump-probe system (see **Methods** and **Supplementary Note-3**). The bottom panel of Fig. 2(b) summarizes the recorded emission spectra under different pump fluences of beam A only ($P_A$, here $P_B$ = 0). The diameter of excitation beam is fixed at *D = 5 μm* in this research. Broad photoluminescence peak is achieved at low pump fluence and a sharp peak emerges when $P_A$ is around *250 nJ/cm$^2$*. The sharp spike slightly shifts to shorter wavelength due to the band filling effect (left panel in Fig. 2(c)). Its intensity increases dramatically and dominates the spectrum at higher pump fluence. Such behaviors are associated with the slope change in log-log plot of light-light curve (left panel in Fig. 2(d)) and a donut beam appears in far field (inset in Fig. 2(d)). All these observations confirm the quasi-BIC microlaser with a threshold of $P_{th}$ = *250 nJ/cm$^2$* in our metasurface. Then the location of excitation beam varies and the corresponding lasing actions are recorded. The good uniformity in both lasing wavelength and threshold is confirmed experimentally (**Supplementary Note-3**). The quasi-BIC laser mostly propagates in-plane along *x*- and *y*-directions (inset of Fig. 2(d)) and has restricted loss. Therefore, BIC metasurface can be an ideal platform to explore the proposed distant interaction.

Then we fix the power of beam A at $P_A$ = *1.2 $P_{th}$* and turn on beam B, which is 30 *μm* (center to center distance, $d_{AB}$) away from beam A along *x*-direction. While two beams are widely separated (~50λ), we can still see the changes in lasing characteristics with the increase of pump power $P_B$. The emission wavelength shifts to shorter wavelength and the emission intensity decreases with the total pump fluence ($P_{total} = P_A + P_B$) (top panel in Fig. 2(b)). When $P_B$ is above *0.6 $P_{th}$*, the lasing mode disappears and only a broad photoluminescence remains. The broad photoluminescence peak above the laser threshold is well preserved over a large power range of $P_B$ from *0.6 $P_{th}$* to *0.85 $P_{th}$* (*1.8 $P_{th}$ < $P_{total}$ < 2.05 $P_{th}$*). With a further increase of $P_B$, the lasing peak re-emerges and a doublet can be clearly seen in the emission spectrum.

The wavelengths and intensity of lasing modes are summarized in right panels of Figs. 2(c, d). The reduction in intensity, disappearance of lasing mode, and mode spitting can be more clearly seen, consistent with the theoretical model of laser self-termination very well. Meanwhile, the separation distance $d_{AB}$ = 30 μm is orders of magnitude larger than the carrier diffusion length (~ 10 nm, see **Supplementary Note 4**) in perovskite film.[36] The long-range interaction between two quasi-BIC microlasers can thus be confirmed for the first time.

Quasi-BIC microlasers are defined by excitation beam at arbitrary locations of metasurface. Then the coupled system can be reconfigured by changing the separation distance between two beams and the maximal coupling distance can be experimentally determined. Here two beams reach the sample simultaneously and their pump fluences are *1.2 $P_{th}$* and *0.8 $P_{th}$*, respectively. We experimentally vary the position of beam B along x-direction and all the results are summarized in Fig. 3(a). Two pump beams are obviously separated when $d_{AB}$ is larger than 9 μm. Under such a situation, strong coupling occurs and a lasing doublet can be observed. With the increase of $d_{AB}$, two lasing modes approach one another and gradually disappear at $d_{AB}$ ≥30 μm. The lasing peak reappears at $d_{AB}$ > 45 μm. It becomes a single mode again and its intensity gradually increases with $d_{AB}$. All these observations are also attributed to the mode interaction. In experiment, the propagation in perovskite film is affected by the material loss and exponentially decay with the distance. Taking the position dependent J into accounted, the eigenvalues have been calculated and plotted as lines in Fig. 3(a). Both the doublet at short distance and the lasing self-termination phenomenon can be reproduced (**Supplementary Note-4**). When the separation distance $d_{AB}$ is above 60 μm, we find that the emission intensity of coupled system saturates at the value of a single microlaser with a pump fluence of *1.2 $P_{th}$*. This shows that the laser self-termination effect is negligibly small at $d_{AB}$ > 60 μm. Then we know that the quasi-BIC microlasers can at least interact each other with a separation distance up to ~ 110λ.

Another important feature between long-range interaction from evanescent coupling is its dynamic controllability. According to Eq. (2), the eigenvalues of hybrid states can also be controlled with the difference in gain coefficients of two quasi-BICs ($\Delta\gamma = (\gamma_a - \gamma_b)/2$. In principle, the gain coefficients $\gamma_a$ and $\gamma_b$ are closely related to the population inversion. For the case of ultrafast excitation, it is essential to note that $\gamma_a$ and $\gamma_b$ are no longer time invariant. Instead, they change as a function of time due to the evolution of population inversion after the ultrafast excitation (**Supplementary Note-5**). As a consequence, even though the fluences of two excitation beams are fixed, the difference in gain coefficient ($\Delta\gamma$) can still be time dependent if they reach the sample at different time. This simply triggers the possibility of dynamically controlling the distant coupling with a delay time between two beams ($\Delta t_{AB} = t_B - t_A$, the negative value means that beam B arrives earlier).

Then we verify this possibility in experiment. Here two beams are separated with a distance of $d_{AB}$ = 30 μm and their fluences are fixed at *1.2 $P_{th}$* and *0.8 $P_{th}$*, respectively. The delay time $\tau_{AB}$ is controlled by a delay line (**Supplementary Note-3**). The emission spectra at different $\Delta t_{AB}$ have been recorded and analyzed in Fig. 3(b). The gain coefficient ($\Delta\gamma$) increases as a function of $\Delta t_{AB}$. According to Eq. (2), the interaction between two modes increases too and the laser self-termination phenomenon occurs. This analysis has been experimentally confirmed with the trend of laser intensity. It decreases with the increase of $\Delta t_{AB}$ at the beginning. Further increase of $\Delta t_{AB}$ can fully suppress the lasing mode at $\Delta t_{AB} = -4.5\ ps$, leading to the well-known lasing death effect. Similarly, the lasing mode reappears at $\Delta t_{AB} \geq 5\ ps$ and an obvious laser doublet can be seen. Therefore, we confirm that the long-range interaction in our metasurface can be dynamically and ultrafast controlled. The corresponding emission spectrum can be switched from single-mode laser emission to photoluminescence and mode splitting within a delay time around 10 $ps$.

**Multiple coupled quasi-BIC microlasers and the ultrafast control**

One BIC metasurface is able to support a series of quasi-BIC microlasers with near-zero detuning at arbitrary locations. This characteristic, associated with the long-range reconfigurable interaction, enables the coupling between multiple nodes and eventually leading to a scalable and reconfigurable network. Due to the limitation of our optical setup, here we take the three-site interaction as examples to illustrate this potential. We first consider a simple co-linear configuration of three-site interaction (Fig. 4(a)). In principle, this configuration can be described with a 3×3 non-Hermitian Hamiltonian (**Supplementary Note-6**), and its non-Hermitian particle-hole symmetry of theoretically allows non-Hermitian zero mode. In experiment, we set the pump fluences of three beams of $P_A=P_B=P_C=1.1P_{th}$ and their separation distances of $d_{AC}=d_{BC}=15\ \mu m$. The coupled system produces three lasing modes (central panel in Fig. 4(b)) when three beams pump the sample simultaneously. Following the theoretical model and previous report, the central mode is the designed zero mode.[37].

In contrast to evanescent coupling between traditional optical cavities, the long interaction distances allow individual control of pump fluences and their relative delay time ($\Delta t_{AB}, \Delta t_{AC}, \Delta t_{BC}$), enabling the precise of zero-mode lasing action. In our experiment, we adjust the time delay between the middle beam (beam C) and the other two beams ($\Delta t_{AB} = 0$ and $\Delta t_{AC} = \Delta t_{BC}$) and explore the lasing behaviors. The corresponding results are summarized in Fig. 4(c). For the case of $\Delta t_{AC} < 0\ ps$, the gain of beam C decreases when the other two beams reach the sample. According to the theoretical model, the system can reach a new status where only zero-mode lase (**Supplementary Note-6**). This is exactly what we have observed experimentally. As shown in bottom panel of Fig. 4(b) and Fig. 4(c), three lasing peaks collapse and only the zero-mode lasing is achieved. The corresponding far-field pattern shows two lobes (43). This is consistent with the non-Hermitian zero-mode theory that π/2 phase jump between adjacent cavities occurs. When beam C pumping the sample later, (e.g. $\Delta t_{AC} > 0$ ), the dominant interaction occurs between A and B. Then only lasing doublet can be seen and matches the theoretical model well (**Supplementary Note-6**). The ultrafast transitions between different states are intrinsically different from conventional approaches. [37]

In additional to controllable zero-mode laser, three-site interaction can also tailor the light emission at lower pump fluences. To demonstrate the flexibility of coupled system in BIC metasurface, we switch the three pump beams to a right isosceles triangle (see Fig. 4(d)). Here the pump fluences of three beams are all below threshold (e.g., $P_A=P_B=P_C=0.9\ P_{th}$) and their diameters remain at $D = 5\mu m$. Their separation distances are $d_{AB}=25~\mu m$ and $d_{AC}=d_{BC}=18~\mu m$. We first explore the case of single beam without considering the interactions between different quasi-BICs. The results are summarized in **Supplementary Note-7**. If the pump fluence is fixed at $0.9P_{th}$, only a broad photoluminescence has been achieved no matter the beam size is 5 μm, 10 μm, 20 μm, and even 50 μm. The situation becomes completely different when three quasi-BICs are formed and their interactions are included. With the control of delay time, we find that one state of the coupled system has been increased significantly to above the laser threshold. In other words, a lasing mode has been achieved at around $\Delta t_{AC} = 0$ ps (see Figs. 4(e) and 4(f)) even though all three quasi-BICs can't lase by themselves. Such kind of single-mode laser emission can only be generated within a short time range of $|\Delta t_{AC} \leq 10~ps|$, enabling ultrafast control of our coupled system as well. Note that the interaction of multiple-site is limited by the delay lines to demonstrate the ultrafast control. The ultralong coupling distance in BIC metasurface certainly ensures the realization of large coupled laser arrays. The controllable long-range interaction and nonlinearity of microlasers make coupling systems based on BIC metasurfaces very promising in large-scale optical computing and reconfigurable photonic neural networks.

**Conclusion**

In summary, we demonstrate that BIC metasurface can function as an ideal platform for the controllable long-range interaction. The uniform lasing wavelength and two-dimensional BIC waveguide greatly increase the interaction distance up to tens of micrometers, which introduces the dynamic control of mode coupling and enables the construction of scalable and reconfigurable photonic circuits. Our concept can be extended to a large array of coupled microlasers and even passive cavities. It is thus able to empower the researches of topological photonics[38, 39], supersymmetry[40, 41], reservoir computing[2, 3], and quantum network[42] with precisely controllable coupling constants. This research represents a new paradigm for scalable coupled systems and shall revolutionize the optical computing and quantum information processing.

## Methods

**Numerical simulation.** The band structures, Q-factors, and the corresponding field pattern were calculated with a finite-element method using commercial software (COMSOL Multiphysics). Periodic boundary conditions are applied in the x- and y- directions to mimic the infinite large periods. Perfectly matched layers are used the z direction to fully absorb the outgoing waves. Optical constants of the active layer and top polymer are obtained from experimental results of Quasi-2D perovskite and ZEP520A resist, which are measured by ellipsometry. The refractive index of glass substrate is set as 1.45.

**Sample Fabrication.** The BIC metasurface is fabricated on a 13 nm ITO coated glass substrate by a combined process of spin-coating of quasi-2D perovskite film and electron beam lithography. The precursor solutions of $N_2F_8$ ($(NMA)_2FA_{n-1}PbBr_{3n+1}$, n=8) was prepared by solving a 25% molar ratio of 1-naphthylmethylamine bromide (NMABr) into a mixture of $HC(NH_2)_2Br$ (FABr) and $PbBr_2$ (1:1 ratio) in DMF at 0.4 M and stirred at 60 °C for 12 hours. The mixtures were then spin-coated on the ITO layer at 5,000 r.p.m. for 30 seconds. During spin coating, 0.3 ml of ethyl acetate was dropped onto the perovskite precursor layer. The substrates were baked on a hotplate at 85 °C for 15 min. Then 100 nm ZEP260A was spin-coated onto the film and patterned with electron-beam lithography (Raith e-LINE). After developing in N50 for 60s, the metasurface was eventually achieved.

**Optical Characterization.** The sample was optically excited by a frequency-doubled Ti:Sapphire laser (400 nm, repetition rate 1 kHz, pulse width 100 fs). The incident laser was first divided into three beams, two of which passed through two delay lines respectively. Three beams were then combined and focused onto the surface of the sample using an objective lens (40X, NA = 0.65). The spatial positions and delay times of three beams are controlled by spatial deviations and delay lines, respectively. The experimental details have been summarized in Supplementary Note-3.

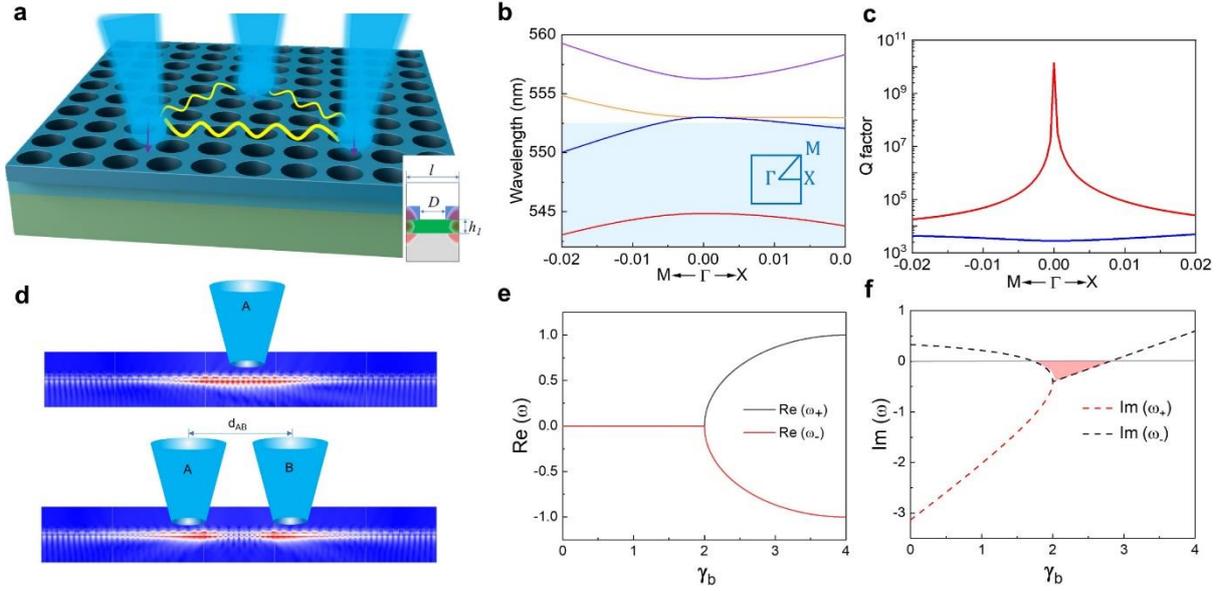

**Figure 1. Interactions between quasi-BIC microlasers.** (**a**) Schematic of the BIC metasurface. Inset shows the near-field mode profile in x-z plane. (**b**) Band structure of the BIC nanostructure. The gain spectrum is marked with a shadowed area. (**c**) Q-factors of resonances within the gain spectrum. (**d**) Schematic of long-range photon propagation (top) and long-range interaction (bottom) via BIC waveguide. (**e**) and (**f**) show the real and imaginary parts of the eigenvalues of two coupled quasi-BIC resonators. For simplicity, we define $\omega_0 = 0$, $\gamma_a = 4$, $J_{ba} = J_{ab} = 1$, $\kappa_a = \kappa_b = 3.4$. The lasing threshold is defined as $Im(\omega) = 0$.

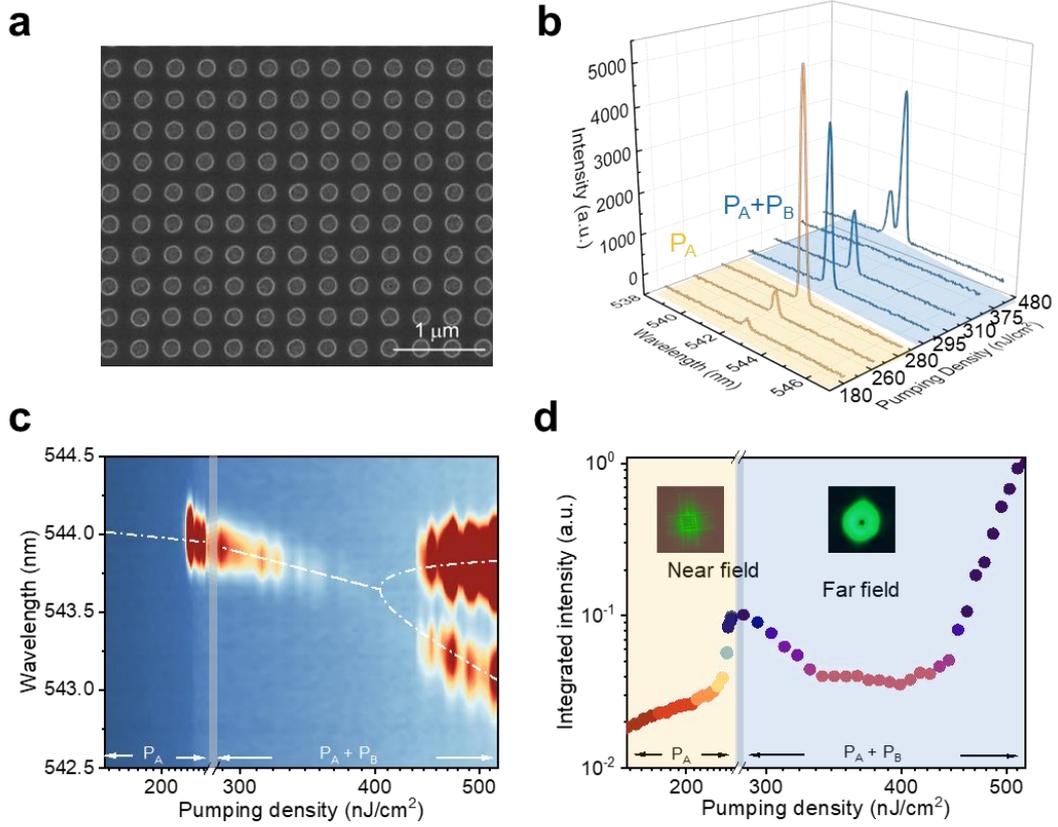

**Figure 2. Distant interaction between two quasi-BIC microlasers.** (**a**) Top-view SEM image of the perovskite BIC metasurface. It is the square-latticed nanohole array in ZEP520A. (**b**) The emission spectra of BIC metasurface at different pump fluences. Bottom panel shows emission spectrum of a single quasi-BIC microlaser. Top panel shows the emission spectra as a function of pump fluence of beam-B. (**c**) and (**d**) are the lasing wavelengths and the integrated output intensity as a function of pump fluences of beam A only (magenta region) and two beams (blue region). Here the fitting parameters are set as $\text{Re}(\omega_0) = 3462$ THz, $\gamma_a = 16$ THz, $0 < \gamma_b < 16$ THz, $J = 4$ THz, $k = 14$ THz. The insets show the near field and far field images of quasi-BIC microlasers. In two beams experiment, the pump fluence of beam-A is fixed at 1.2 $P_{th}$.

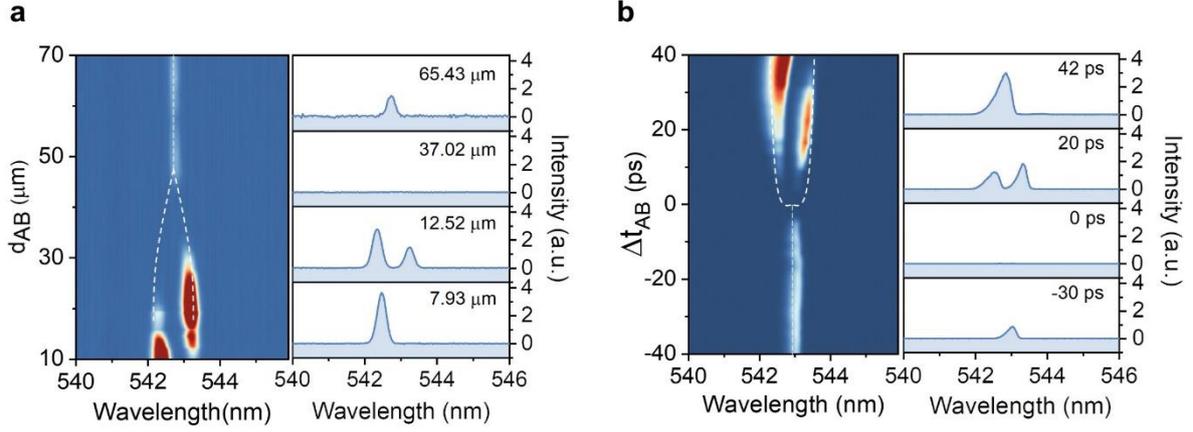

**Figure 3. Control of distant coupling between two quasi-BICs.** (**a**). Left: two-dimensional colourmap of emission spectrum as a function of separation distance between two beams. Right: Emission spectra at several particular distances such as $d_{AB}$ = 7.93 *μm*, 12.52 *μm*, 37.02 *μm*, 65.43 *μm*, respectively. Complete suppression of lasing modes can be observed under two types of controls. (**b**) Left: two-dimensional colourmap of emission spectrum as a function of time delay between two excitation beams. Dashed line and solid line represent the theoretical fitting following coupled mode theory. Right: Emission spectra at particular time delays, i.e., $\Delta t_{AB}$ =- 30 *ps*, 0 *ps*, 20 *ps*, 42 *ps*, respectively.

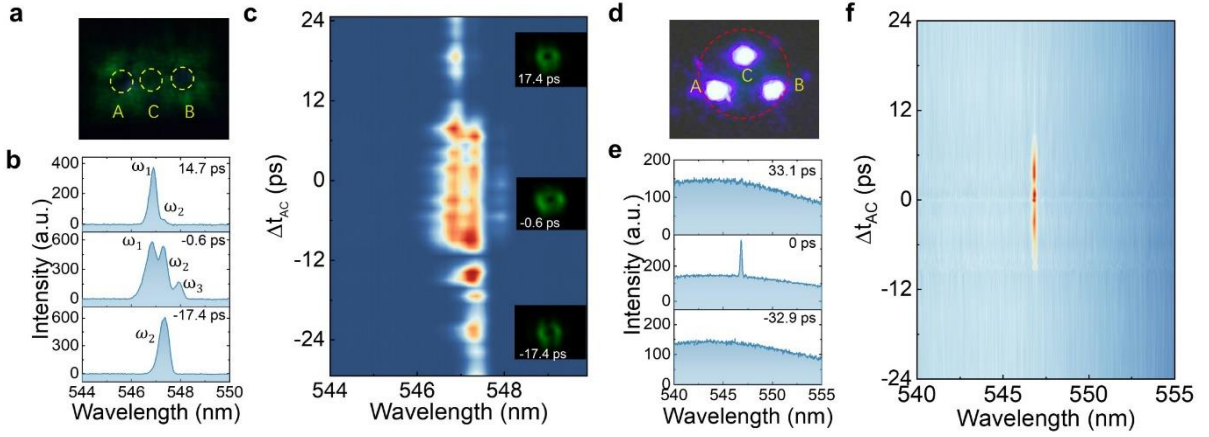

**Figure 4. Dynamical control of three-site coupled systems.** (**a**) Near field image of zero-mode lasing pumped by three collinear beams. The pump fluences here are set as $P_A=P_B=P_C=1.1P_{th}$, the delay times $\Delta t_{AB}=0$, $d_{AC}=d_{BC}$=15 μm. (**b**) The recorded emission spectra at $\Delta t_{AC}$= -17.4 ps, -0.6 ps, and 17.4 ps. (**c**) The evolution of emission spectrum with the increase of $\Delta t_{AC}$. (**d**) The pump configuration for controllable optical gain. Here the pump fluences are $P_A=P_B=P_C=0.9P_{th}$. The delay time of $\Delta t_{AB}=0$, and the separation distances are $d_{AC}=d_{BC}$=18 μm, $d_{AB}$ =25 μm. (**e**) Typical emission spectra at $\Delta t_{AC}$= -32.9 ps, 0 ps, and 33.1 ps. (**f**) summarizes the evolution of emission spectrum as a function of $\Delta t_{AC}$.

**Acknowledgements**
The authors acknowledge support by National Key Research and Development Program of China (Grant Nos. 2021YFA1400802 and 2022YFA1404700), National Natural Science Foundation of China (Grant Nos. 6233000076, 12334016, 11934012, 12025402, 62125501 and 62305084), Guangdong Basic and Applied Basic Research Foundation 2023A1515011746, Shenzhen Fundamental Research Projects (Grant Nos. GXWD 2022081714551), Shenzhen Science and Technology Program (Grant Nos. JCYJ20230807094401004).

**Author contributions**
Q.S., C.H. conceived the idea and supervised the research. H.T., C.H., Y.W., X.J., S.X., J.H. and Q.S. did the design. H.T., Y.W. and X.J. fabricated the samples. C.H. and H.T. performed the experimental measurements. Q.S., C.H. and H.T. analysed the results. All the authors discussed the contents and prepared the manuscript.

**Competing interests**
The authors declare no competing interests.